\documentclass[aps,preprint]{revtex4}%
\usepackage{amsfonts}
\usepackage{amsmath}
\usepackage{amssymb}
\usepackage{graphicx}%
\providecommand{\U}[1]{\protect\rule{.1in}{.1in}}
\newtheorem{theorem}{Theorem}

\newtheorem{lemma}[theorem]{Lemma}

\begin{document}
\title{Mathematical analysis of long-time behavior of magnetized fluid instabilities with shear flow}
\author{Youngmin Oh}
\affiliation{Beijing Computational Science Research Center, Building 9, East Zone, ZPark
II, No.10 East Xibeiwang Road, Haidian District, Beijing 100193, China}
\email{youngminoh19850329@csrc.ac.cn}
\author{Gunsu S. Yun}
\affiliation{Department of Physics, Pohang University of Science and Technology, Pohang,
Gyeongbuk 37673, Republic of Korea}
\email{gunsu@postech.ac.kr}
\author{Hyung Ju Hwang}
\affiliation{Department of Mathematics, Pohang University of Science and Technology,
Pohang, Gyeongbuk 37673, Republic of Korea}
\email{(corresponding author) hjhwang@postech.ac.kr}
\begin{abstract}
We study a complex Ginzburg-Landau (GL) type model related to fluid instabilities in the boundary of magnetized toroidal plasmas (called edge-localized modes) with a prescribed shear flow  on the Neumann
boundary condition. We obtain the following universal results for the model in
a one-dimensional interval. First, if the shear is weak, there is a unique
linearly stable steady-state perturbed from the nonzero constant steady-state
corresponding to the zero shear case. Second, if the shear is strong, there is
no plausible steady-state except the trivial zero solution in the interval.
With the help of these results and the existence of global attractors, we can
dramatically reduce the number of cases for the long-time behavior of a
solution in the model.
\end{abstract}
\maketitle

\renewcommand{\theequation}{\arabic{equation}}

\section{Introduction\label{sec1}}

Relaxation phenomena are common in nature \cite{rp1,rp2}, the most notable
examples in magnetized plasmas being the explosive flares on the surface of
the Sun. In toroidally confined plasma (e.g., tokamak), semi-periodic
explosive bursts occur at the so-called $H$-mode plasma boundary. The $H$-mode
corresponds to a state of highly improved confinement of heat and particles,
and such a state is routinely obtained when the heating applied to the
confined plasma exceeds a particular threshold. The lower confinement state is
called $L$-mode. During the transition from $L$- to $H$-mode, a transport
barrier spontaneously builds up due to high $E\times B$ flow shear at the
plasma edge, which considerably reduces heat and particle transports
\cite{eb3,eb5, eb2,eb4,eb1,eb6}.

However, this edge barrier is linearly unstable to a class of helical
filamentary eigenmodes called edge localized modes (ELMs). Eventually, the
barrier relaxes violently with a rapid expulsion of particles and heat,
presumably triggered by the burst of the ELM filament. For this reason, the
barrier relaxation is commonly called ELM crash although the recent
experiments and simulations suggested that the relaxation is triggered by
solitary perturbations rather than by ELM eigenmodes \cite{krebs2013nonlinear,
lee2017solitary, wenninger2012solitary}.

The ELM crash should be avoided because the significant heat and particle flux
can damage the plasma-facing walls albeit small beneficial effects such as
removal of impurity particles from the plasma. It is crucial to understand the
ELM dynamics to prevent or mitigate the explosive barrier relaxation. For a
possible explanation of the nonlinear relaxation oscillations,
phenomenological models for several types of ELMs have been proposed
\cite{beyer2005nonlinear, m1, leconte2016ginzburg, m2}.

Nevertheless, as far as we know, the linear stability analysis has dominated
most theoretical studies on ELMs \cite{ls}, and there has been little work
related to the dynamic behavior of ELMs (nonlinear oscillation). Accordingly,
at present, the nonlinear mechanism is not fully clarified, and more thorough
mathematical analysis is required. Especially, we emphasize that there is no
precise explanation why the nonlinear oscillations exist in the $H$-mode
plasma boundary. To solve it, recently, a Ginzburg-Landau type model was
proposed with a prescribed mean shear flow (MSF) \cite{leconte2016ginzburg}
based on the critical gradient model \cite{beyer2005nonlinear}.

\subsection{Main results}
We employ the following Ginzburg-Landau (GL) type equation for the pressure perturbation
 $P\left(  t,x\right)  \in\mathbb{C}$ in the slab geometry with the space domain
$\Omega=\left[  -1,1\right]  \subset\mathbb{R}$ developed in
\cite{leconte2016ginzburg}:
\begin{align}
\partial_{t}P\left(  t,x\right)  -\mu\partial_{x}^{2} P\left(  t,x\right)  +\gamma
_{N}\left\vert P\left(  t,x\right)  \right\vert ^{2}P\left(  t,x\right)
-\left(  \gamma_{L}-iAV(x)\right)  P\left(  t,x\right)   &  =0,\label{main1}\\
\partial_{x}P\left(  t,\pm1\right)   &  =0,\nonumber
\end{align}where $\partial_{x}^{k} f =\frac{\partial^k f}{\partial x^k}$ and $\partial_{t}^{k} f =\frac{\partial^k f}{\partial t^k}$ for a given function $f$ with $0\leq k \in \mathbb{Z}$.
Here, $A>0$ is the shear flow strength including the effect of finite poloidal
wavenumber, $V$ is the normalized prescribed mean-sheared function,
$\gamma_{L}>0$ is the linear growth rate, $\mu>0$ is the cross-field turbulent
heat diffusivity, and $\gamma_{N}>0$ is the nonlinear damping. 
Notice that $P\left(  x\right)  $ with $\left\vert P\right\vert :=\left(
\gamma_{L}/\gamma_{N}\right)  ^{1/2}$ is the unique nonzero constant
steady-state of (\ref{main1}) for $A=0.$ If we denote $P\left(  t,x\right)
=R\left(  t,x\right)  \exp\left(  i\theta\left(  t,x\right)  \right)  $, then
we can rewrite (\ref{main1}) as%
\begin{align}
R_{t}  &  =\mu\partial_{x}^{2}R-\mu R\left\vert \partial_{x}\theta\right\vert
^{2}+\gamma_{L}R-\gamma_{N}R^{3},\label{m2-1}\\
R\partial_{t}\theta &  =\mu R\partial_{x}^{2}\theta+2\mu\partial_{x}%
R\partial_{x}\theta-AV\left(  x\right)  R. \label{m2-2}%
\end{align}

Substantial theoretical results exist for the GL equation because of its
recurring importance in various subjects of physics. For an extensive review
on GL equations with constant complex coefficients, refer to \cite{r1}. To our
knowledge, however, there are few theoretical results for the GL equation with
variable coefficients despite its potential. Nevertheless, some consequences
with constant coefficients inspired us as follows. 1) a non-zero constant
steady-state of the real GL equation is a unique linearly stable steady-state
in a convex domain for the Neumann boundary condition
\cite{jimbo1994stability}; 2) there exists a global attractor of the CGL
equation \cite{li2014global,w5}. Inspired by them, we can show the
following\ mathematical statement (see Appendix for the notations which are used in the statement). 
\begin{theorem}
\label{t1}Let $0\not \equiv V\left(  x\right)  \in L^{\infty}\left(
\Omega\right)  \ $satisfy $V\left(  -x\right)  =-V\left(  x\right)  $ in
$-1<x<1$ and $V\left(  x_{1}\right)  \leq V\left(  x_{2}\right)  $ if
$x_{1}<x_{2}$ almost everywhere in $\Omega=\left[  -1,1\right]  \subset
\mathbb{R}.$ (a): \textit{For sufficiently small }$A\geq0,$\textit{ there
exists a unique linearly stable steady-state }$P_{A}\left(  x\right)
=R_{A}\left(  x\right)  \exp\left(  i\theta_{A}\left(  x\right)  \right)  \in
H_{n}^{2}\left(  \Omega\right)  $\textit{ of (\ref{main1}) such that }%
$R_{A}\left(  -x\right)  =R_{A}\left(  x\right)  $\textit{ and}%
\[
\lim_{A\rightarrow0}R_{A}(x)=\left(  \gamma_{L}/\gamma_{N}\right)  ^{1/2}.
\]
\textit{Besides, the first eigenvalues of linearly unstable steady-states for
}$A=0$\textit{ are bounded below; (b): For sufficiently large }$A>0,$\textit{
there does not exist nonzero steady-state }$P\left(  x\right)  =R\left(
x\right)  \exp\left(  i\theta\left(  x\right)  \right)  $\textit{ of
(\ref{main1}) such that }$R\left(  -x\right)  =R\left(  x\right)  $\textit{,
}$R^{\prime}\left(  x\right)  \leq0$\textit{ if }$-1<x<0$\textit{ and
}$R^{\prime}\left(  x\right)  \geq0$\textit{ if }$0<x<1.$
\end{theorem}

\begin{figure}
[ptb]
\begin{center}
\includegraphics[
width=\textwidth,
keepaspectratio
]%
{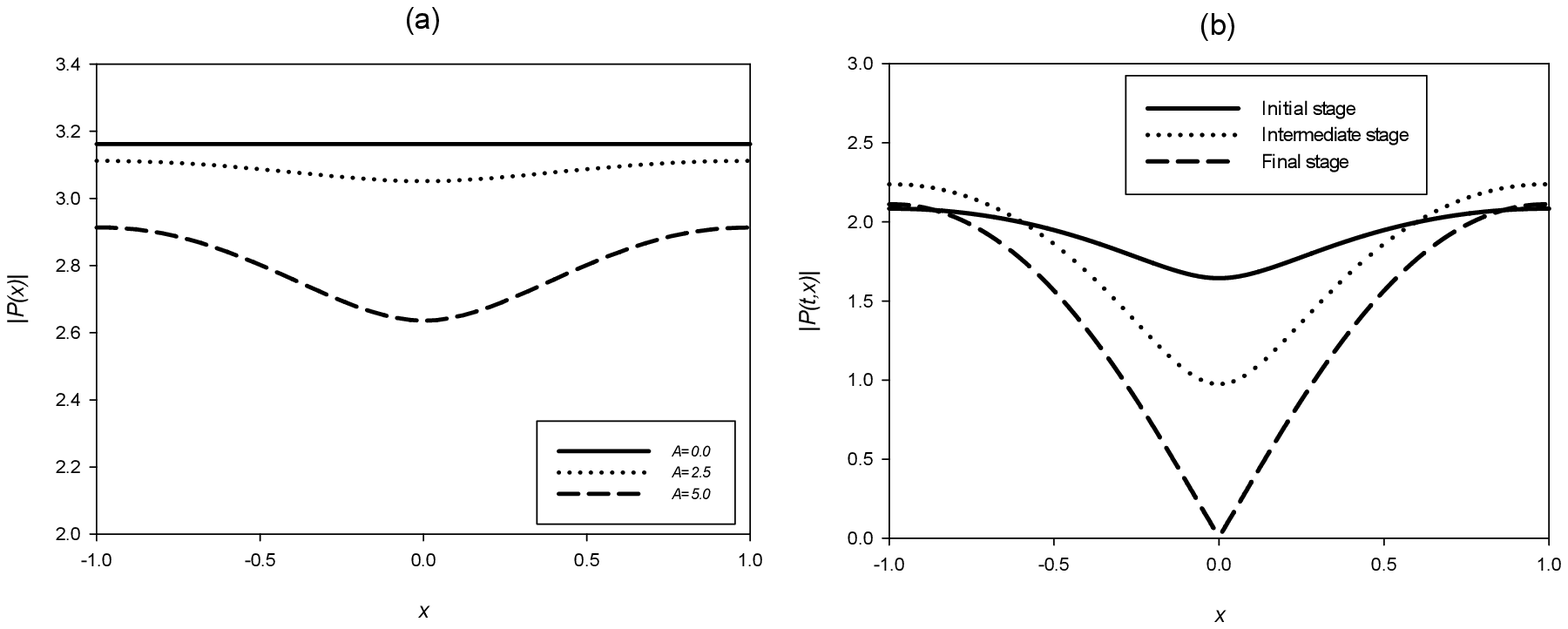}%
\caption{(a) The magnitude $\left\vert P\left(  x\right)  \right\vert $\ of a
steady-state $P\left(  x\right)  $ to (\ref{main1}) for $A$ with $\mu
=\gamma_{N}=1,$ $\gamma_{L}=10,$ and $V\left(  x\right)  =\tanh(25x).$ (b) The
time behavior of the magnitude $\left\vert P\left(  t,x\right)  \right\vert $
of a solution $P\left(  t,x\right)  $ to (\ref{main1}) with $\mu=\gamma
_{N}=1,$ $\gamma_{L}=10,$ $A=10,$ and $V\left(  x\right)  =\tanh\left(
25x\right)  $ (the solid line: an initial stage; the dotted line: an
intermediate stage; the dashed line: a final stage). The oscillation repeats
these stages successively. $R=\left\vert P\right\vert $ in (a)-(b) satisfies
$R^{\prime}\leq0$ in $-1<x<0,$ and $R^{\prime}\geq0$ in $0<x<1.$}%
\label{fig1}%
\end{center}
\end{figure}

Theorem \ref{t1} $(a)$ gives a clue that the behavior of solutions is
independent of initial condition. Indeed, it is possible to show that a
numerical solution $P\left(  t,x\right)  $ for the complex GL equation
(\ref{main1}) converges to the nonconstant steady-state for weak shear $0\leq
A<<1$ (FIG. \ref{fig1} (a)). Therefore, we predict that only the strong
shear leads to ELM crash.

\begin{figure}
[ptb]
\begin{center}
\includegraphics[
width=\textwidth,
keepaspectratio
]%
{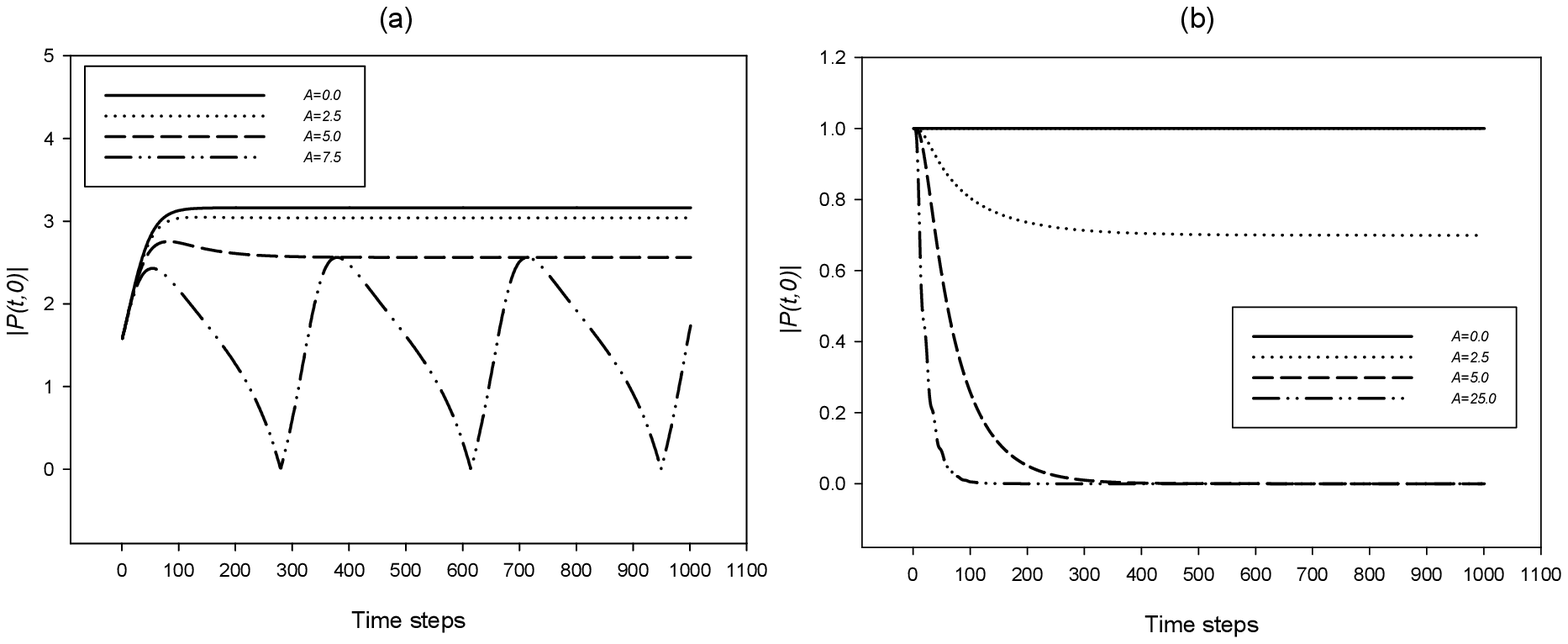}%
\caption{The evolution of the magnitude $|P(t,0)|$ of a solution $P(t,x)$ to
(\ref{main1}) for $A$ with (a) $\mu=\gamma_{N}=1,$ $\gamma_{L}=10,$ $V\left(
x\right)  =\tanh\left(  25x\right)  ,$ and initial condition $P\left(
0,x\right)  =\frac{\left(  \gamma_{L}/\gamma_{N}\right)  ^{1/2}}{2};$ (b)
$\mu=\gamma_{N}=\gamma_{L}=1,$ $V\left(  x\right)  =\tanh\left(  25x\right)
,\ $and initial condition $P\left(  0,x\right)  =\left(  \gamma_{L}/\gamma
_{N}\right)  ^{1/2}.$ The difference of $\gamma_{L}$ between (a) and (b) shows
the different long-time behavior for large $A$ ((a): nonlinear oscillation;
(b): convergence to $0$).}%
\label{fig2}%
\end{center}
\end{figure}

Theorem \ref{t1} $(b)$ provides two possibilities; 1) there are only two types
of the long-time behavior of $P\left(  t,x\right)  $ for large $A$: nonlinear
oscillation or convergence to $0$ (FIG. \ref{fig2} (a)-(b)); 2) the
long-time behavior of $P\left(  t,x\right)  $ completely depends on the linear
stability of the zero solution. These show that the occurrence of
quasiperiodic ELM\ crash may depend on the parameters of the complex GL
equation since the stability of the zero solution is unclear.

The detailed proof of Theorem \ref{t1} is provided in Appendix. The main
reason why we consider the conditions for a steady-state in Theorem \ref{t1}
$(a)$-$(b)$ is that numerical simulations satisfy the conditions for $R$ in
Theorem \ref{t1} $(b)$ (FIG. \ref{fig1} (a)). We point out the existence of
global attractors in $H^{1}\left(  \Omega\right)  $ so that a solution of
(\ref{main1}) is uniformly bounded in $L^{\infty}\left(  \Omega\right)  $. It
suffices to show that $\left\Vert P\right\Vert _{H^{1}}^{2}+\left\Vert
P\right\Vert _{4}^{4}$ and $\left\Vert P\right\Vert _{H^{2}}^{2}+\left\Vert
P\right\Vert _{6}^{6}$ are uniformly bounded in time to show the existence of
the global attractor in $H^{1}\left(  \Omega\right)  $. For detailed reason,
please see \cite{li2014global}.

\begin{lemma}
\label{mlemma1} Let $0\not \equiv V\left(  x\right)  \in H_{n}^{1}\left(
\Omega\right)  $. Then there exists a global attractor of (\ref{main1}) in
$H^{1}\left(  \Omega\right)  $ for $A\geq0.$
\end{lemma}

In Section \ref{nv}, we compare Theorem \ref{t1} with numerical results. We
summary and conclude our results in Section \ref{con}.%

\begin{figure}
[ptb]
\begin{center}
\includegraphics[
width=\textwidth,
keepaspectratio
]%
{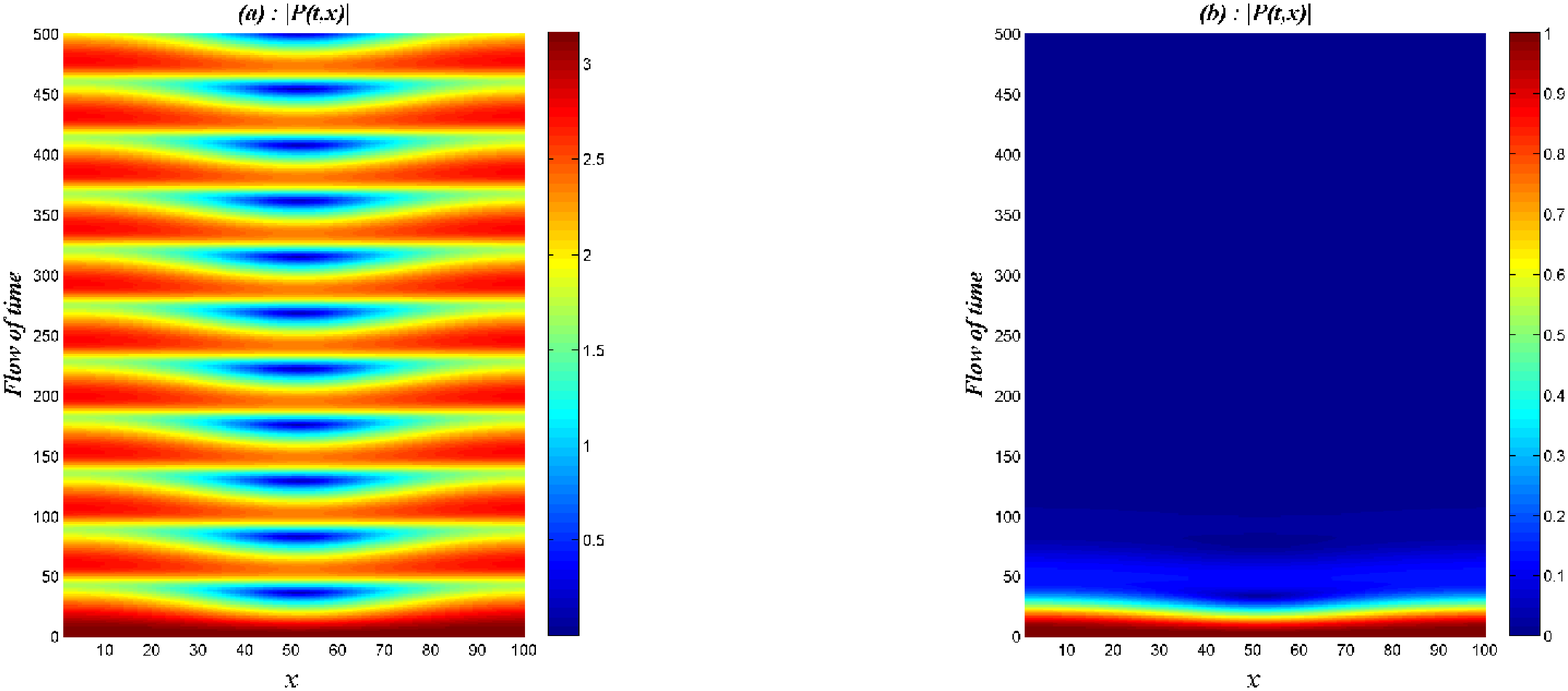}%
\caption{The evolution of the magnitude $\left\vert P\left(  t,x\right)
\right\vert $ (color bar) of a solution $P\left(  t,x\right)  $ to
(\ref{main1}) for $A=10$ with (a) $\mu=\gamma_{N}=1,$ $\gamma_{L}=10,$
$V\left(  x\right)  =\tanh\left(  25x\right)  ,$ and initial condition
$P\left(  0,x\right)  =\left(  \gamma_{L}/\gamma_{N}\right)  ^{1/2};$ (b)
$\mu=\gamma_{N}=\gamma_{L}=1,$ $V\left(  x\right)  =\tanh\left(  25x\right)
,\ $and initial condition $P\left(  0,x\right)  =\left(  \gamma_{L}/\gamma
_{N}\right)  ^{1/2}.$ The difference of $\gamma_{L}$ leads to
the different long-time behavior for large $A$ ((a): nonlinear oscillation;
(b): convergence to $0$)$.$}%
\label{fig3}%
\end{center}
\end{figure}

\section{Numerical Verifications \label{nv}}

We conduct numerical simulations for the pressure perturbation
 to support Theorem
\ref{t1}. We impose $V\left(  x\right)  =\tanh\left(  25x\right)  \ $for
numerical verifications. It is possible to confirm that the small shear leads
to a steady-state in FIG. \ref{fig1} (a). It is observed that the magnitude
$|P(x)|=R(x)$ of all steady-states $P(x)=R(x)\exp(i\theta)$ in FIG.
\ref{fig1} (a) decreases in $x\in\left[  -1,0\right]  $ and increases in
$x\in\left[  0,1\right]  $. This attribute in the profiles of steady-states is
observed for a wide range of parameters of the complex GL equation. The shear
acts as a force to keep down the magnitude $R(x)$ such that the force attains a
maximum at the center $x=0$. Although we do not provide a figure of the
evolution of $P\left(  t,x\right)  $ in $t,$ it was also possible to check
that solutions converge to a steady-state for various initial conditions with
a given small $A,$ so the long-time behavior of $P\left(  t,x\right)  $ for
the weak shear is independent of initial conditions. These results are
consistent with Theorem \ref{t1} $(a)$.

If the shear is large and GL parameters are suitable, then $|P\left(
t,x\right)  |$ oscillates nonlinearly and never converges to a steady-state.
Indeed, FIG. \ref{fig1} (b) displays the nonlinear oscillation. The solid
line represents $|P\left(  t,x\right)  |$ when $|P\left(  t,0\right)  |$ peaks
during the oscillation. After that, $|P\left(  t,0\right)  |$ drops to $0,$ so
$|P\left(  t,x\right)  |$ reaches the dashed line finally. After the dashed
line, $|P\left(  t,x\right)  |$ returns sharply to the solid line again. The
dotted line shows an intermediate stage of the oscillation. $|P\left(
t,x\right)  |$ repeats this periodic oscillation without dissipation or
blow-up. Moreover, many numerical results showed that $|P\left(  t,x\right)
|$ decreases in $x\in\left[  -1,0\right]  $ and increases in $x\in\left[
0,1\right]  $ for any time $t>0$ during the nonlinear oscillations$.$ These
oscillations are related to quasiperiodic ELM crash \cite{leconte2016ginzburg}.

FIG. \ref{fig2} (a) shows that the change of the time behavior of $|P\left(
t,0\right)  |$ for $A.$ $|P\left(  t,0\right)  |$ converges to a constant for
small $A$ ($0.0,$ $2.5,$ and $5.0$ in FIG. \ref{fig2} (a)), meaning that
$|P\left(  t,x\right)  |$ converges to a steady-state. However, in the case of
$A=7.5,$ it is shown that $|P\left(  t,0\right)  |$ oscillates without
dissipation. Theorem \ref{t1} $(a)$ explains this bifurcation to the nonlinear
oscillation from the steady-state because it is hard to expect steady-states
except the zero solution for large $A$.

However, we emphasize that Theorem \ref{t1} $(b)$ provides a possibility that
a solution may converge to $0$. Indeed, FIG. \ref{fig2} (b) shows the
convergence to the zero solution in $t$ with the change of $\gamma_{L}$ to $1$
from $10$ for large $A$ ($25.0$ in FIG. \ref{fig2} (b)). Therefore, the long
time behaviors between $\gamma_{L}=1$ and $\gamma_{L}=10$ are completely
different for large $A$. FIG. \ref{fig3} (a)-(b) represent the time behavior
of $|P(t,x)|$ with $A=10$ for $\gamma_{L}=1\ $and $10$ respectively. We also
tested many cases by changing parameters and concluded that there are only two
cases, that is, the nonlinear oscillation and the convergence to the zero solution.

It is important to understand the individual components of (\ref{m2-1}) to
evaluate the relevance of our analysis to the experimental observed ELM
dynamics. There are the dissipation term $\mu R^{\prime\prime}$ which has a
role for flattening $R$, the linear growth term $\gamma_{L}R,$ nonlinear decay
term $\gamma_{N}R^{3}$, and the term $\left\vert \theta^{\prime}\right\vert
^{2}R$ related to the shear although the shear affects $\theta^{\prime}$
implicitly. Without the shear, the magnitude $R(t,x)$ and the phase
$\theta^{\prime}(t,x)$ of the pressure perturbation
 $P\left(  t,x\right)  =R\left(
t,x\right)  \exp(i\theta\left(  t,x\right)  )$ converge to $\left(  \gamma
_{L}/\gamma_{N}\right)  ^{1/2}$ and $0$ respectively as $t\rightarrow\infty$,
but with nonzero $A>0,$ $\left\vert \theta^{\prime}\right\vert $ is nonzero,
so the term $R\left\vert \theta^{\prime}\right\vert $ has a role to press $R$
perpetually$.$ Besides, during the evolution, $\left\vert \theta^{\prime
}\right\vert $ is increasing in $-1<x<0,$ decreasing in $0<x<1$, and maximum
at $x=0$ at any time $t$ for suitable initial conditions$.$ Consequently, the
nonzero shear $\left(  A>0\right)  $ acts like a force to press $R$ such that
it attains its maximum at the center $\left(  x=0\right)  $. This is why the
profile of $R\left(  t,x\right)  $ is minimum at $x=0.$ For large shear
($A>>1$), the pressing force is dominant for $R>0$, so $R\left(  t,0\right)  $
approaches $0.$ During this procedure, the change of the slope of $R$ in a
neighborhood of $x=0$ is so large that the term $\mu R^{\prime\prime}$
strongly affects the profile of $R$ as a restoration force with $\gamma_{L}R$.
Therefore, the terms $\mu R^{\prime\prime}$ and $\gamma_{L}R$ acting as the
restoring forces and the term $R\left\vert \theta^{\prime}\right\vert $ acting
as the pressing force compete with each other. Here, we ignore $\gamma
_{N}R^{3}$ due to $\gamma_{N}R^{3}<<\gamma_{L}R\ $if $R<<1$. $\gamma_{N}R^{3}$
has a role to prevent the blowing-up of $R$. If $\gamma_{L}$ is weak, the
restoring forces are insufficient to restore completely, so $R\left(
t,x\right)  $ converges to $0.$ Conversely, if $\gamma_{L}$ is strong,
$R\left(  t,x\right)  $ has a tendency to restore completely, so $R\left(
t,x\right)  $ oscillates periodically.

The numerical results (in particular, FIG. \ref{fig2} (a)) along with
Theorem \ref{t1} provide a coherent picture on the ELM dynamics observed in
the KSTAR tokamak \cite{lee2017solitary, yun2011two, mp3}. The quasi-steady
ELM state \cite{lee2017solitary,yun2011two} may correspond to the situation
where the shear flow is not sufficiently developed and below the bifurcation
threshold. Furthermore, if the shear parameter $A$ is allowed to evolve in
time, the abrupt transition from the quasi-steady state to the crash state
\cite{lee2017solitary,yun2011two} may correspond to the situation where the
shear flow gradually increases and exceeds the bifurcation threshold, which is
plausible in the boundary of $H$-mode plasmas \cite{eb3}. This scenario would
lead to quasi-periodic ELM oscillations (development of ELM and its crash),
the common situation for the conventional $H$-mode plasmas. On the other hand,
the suppression of the burst of ELM filaments by external magnetic
perturbations \cite{mp3} may be explained by the reduction of the flow due to
the magnetic perturbation (i.e., weakening of the parameter $A$) below the
bifurcation threshold.

To conclude, the shear acts like a force to press the magnitude of the
pressure perturbation, especially, strongly in a neighborhood of the center $x=0,$
so the sharp change around $x=0$ occurs if the shear is large. If $\gamma_{L}$
is large compared to $\mu>0$, the dissipation term $\mu R^{\prime\prime}$ and
the linear growth term $\gamma_{L}R$ completely restore $R$ to the original
state, so nonlinear oscillations occur. If the restoring effect by $\mu
R^{\prime\prime}$ and $\gamma_{L}R$ is insufficient, $R$ converges to $0$ as
$t\rightarrow\infty.$

\section{Conclusion \label{con}}

In summary, we obtained theoretical results to dramatically reduce the number
of cases of the long time behavior of $P\left(  t,x\right)  $ in (\ref{main1})
which is a Ginzburg-Landau type model for ELMs in the one-dimensional case
$\Omega=\left[  -1,1\right]  $. Besides, we also confirmed that these are
consistent with the numerical results.

For the weak shear ($A<<1$), it is reasonable that solutions of (\ref{main1})
converge to a steady-state which is obtained in Theorem \ref{t1} $(a)$. For
the strong shear ($A>>1$), it makes sense that solutions are to either
converge to the zero solution or oscillate nonlinearly by Theorem \ref{t1}
$(b)$. Besides, we confirmed that our results coincide with numerical results.

From these theoretical results, we believe that the model (\ref{main1}) for
ELMs is quite reliable and useful for better understanding of the relaxation
behavior which occurs in nature.

\section*{Acknowledgement}

Hyung Ju Hwang was partly supported by the Basic Science Research Program
through the National Research Foundation of Korea (NRF) (2015R1A2A2A0100251).
Gunsu S. Yun was partially supported by the National Research Foundation of
Korea under grant No. NRF-2014M1A7A1A03029881 and by Asia-Pacific Center for
Theoretical Physics. \renewcommand{\theequation}{A.\arabic{equation}}

\section*{Appendix}

In this appendix, we prove Theorem \ref{t1}. For convenience, we set $\mu=\gamma_{L}=\gamma
_{N}=1$ in (\ref{main1}). We will use the following function spaces:
\begin{align}
L^{p}\left(  \Omega\right)&= \left\{ f(x) : \left\Vert f \right\Vert _{p}=\left\Vert f\right\Vert _{L^{p}}<\infty \right\},\nonumber\\ 
H^{k}\left(  \Omega\right)&=\left\{ f(x) : \left\Vert f \right\Vert _{H^k} <\infty \right\},\nonumber\\ 
H_{n}^{k}\left(  \Omega\right)&=\left\{  f\left(  x\right)  \in H^{k}\left(  \Omega\right)  :\partial_{x}f\left(  \pm1\right)  =0\right\},\nonumber\\ 
H_{0}^{k}\left(  \Omega\right)&=\left\{  f\left(  x\right)  \in
H^{k}\left(  \Omega\right)  :f\left(  \pm1\right)  =0\right\},\nonumber
\end{align} with $1\leq p\in\mathbb{R}$ and $0\leq k \in \mathbb{Z}$, where  
\begin{align}
\left\Vert f \right\Vert _{p}=\left( \int_{\Omega} |f(x)|^p dx\right)^{1/p}, \ \left\Vert f \right\Vert _{H^k}=\left( \sum_{j=0}^{k} \left\Vert \frac{d^j f}{dx^j} \right\Vert _{2}^2\right)^{1/2}. \nonumber
\end{align}
Finally, we use $C>0$ to denote a generic constant which is independent of
parameters which we mainly consider in this paper. 
\subsection*{\label{ia}Proof of Theorem \ref{t1} $(a)$}

In the case of $A=0,$ we already know that $R\equiv1\ $and $\theta=c$ satisfy
a solution of (\ref{m2-1})-(\ref{m2-2}) for any constant $c.$ Without loss of
generality, we assume that $\theta\left(  0\right)  =0.$ We consider the
perturbation%
\[
R=1+r,\text{ }\theta=\theta+0,
\]
so that we can express (\ref{m2-1}) as%
\begin{align}
\partial_{t}r-\partial_{x}^{2}r+2r  &  =-3r^{2}-r^{3}-\left(  1+r\right)
w^{2},\label{m3-1}\\
\left(  1+r\right)  \partial_{t}w-\left(  1+r\right)  \partial_{x}w  &
=2w\partial_{x}r-AV\left(  x\right)  \left(  1+r\right)  , \label{m3-2}%
\end{align}
where $w=\partial_{x}\theta.$

From now, we consider that both $r\ $and $\theta$ are independent of $t.$ Then
we can represent (\ref{m3-2}) as%
\begin{equation}
R\left(  x\right)  ^{2}w\left(  x\right)  =A\int_{-1}^{x}V\left(  x\right)
R\left(  x\right)  ^{2}ds. \label{m3-3}%
\end{equation}
Due to $w\left(  \pm1\right)  =0,$ (\ref{m3-3}) should satisfy the following
compatibility condition for the Neumann boundary conditions as
\begin{equation}
\int_{-1}^{1}V\left(  x\right)  R\left(  x\right)  ^{2}ds=0\text{ for }A\neq0.
\label{m3-4}%
\end{equation}
Note that (\ref{m3-4}) can be possible if $R\left(  x\right)  $ is even.
Moreover, if $R\left(  x\right)  $ is even, $w\left(  x\right)  $ should be
even. Indeed, we can derive $w\left(  x\right)  =w\left(  -x\right)  $ if
$R\left(  x\right)  $ is even since
\begin{align*}
R^{2}\left(  x\right)  w\left(  x\right)   &  =\int_{-1}^{1}AV\left(
s\right)  R^{2}\left(  s\right)  ds-\int_{x}^{1}AV\left(  s\right)
R^{2}\left(  s\right)  ds =-\int_{x}^{1}AV\left(  s\right)  R^{2}\left(
s\right)  ds,\\
R^{2}\left(  x\right)  w\left(  -x\right)   &  =\int_{-1}^{-x}AV\left(
s\right)  R^{2}\left(  s\right)  ds=-\int_{x}^{1}AV\left(  s\right)
R^{2}\left(  s\right)  ds,
\end{align*}
for any $x\in\Omega.$ Then we can also deduce that $\theta\left(  -x\right)
=-\theta\left(  x\right)  $ due to $\theta\left(  0\right)  =0.$ We will show
the existence and uniqueness of $r$ and $w$ which satisfy (\ref{m3-1}%
)-(\ref{m3-2}) if $r$ and $w$ are sufficiently small in norms on $H^{2}\left(
\Omega\right)  $ and $H_{0}^{1}\left(  \Omega\right)  $ respectively.

We first consider the following linear problem.
\begin{align}
-\partial_{x}^{2}r+2r  &  =f,\label{e1}\\
-\partial_{x}w  &  =g. \label{e2}%
\end{align}
where $f$ and $g$ are given as $f\in L^{2}\left(  \Omega\right)  $ and $g\in
L^{2}\left(  \Omega\right)  $ such that $\int_{-1}^{1}g\left(  x\right)
dx=0.$ Then there exist unique solutions $r\in H_{n}^{2}\left(  \Omega\right)
$ and $w\in H_{0}^{1}\left(  \Omega\right)  $ of (\ref{e1})-(\ref{e2}) such
that there is a constant $C>0$ such that
\[
\left\Vert r\right\Vert _{H^{2}}\leq C\left\Vert f\right\Vert _{2}^{2},\text{
}\left\Vert w\right\Vert _{H^{1}}^{2}\leq C\left\Vert g\right\Vert _{2}^{2}.
\]

Now we consider the following problem%
\begin{align}
-\Delta r+2r  &  =f\left(  q,\omega\right)  ,\label{n1}\\
-\partial_{x}w  &  =g\left(  q,\omega\right)  , \label{n2}%
\end{align}
with
\begin{align*}
f\left(  q,\omega\right)   &  =-3q^{2}-q^{3}-\left(  1+q\right)  \omega^{2},\\
g\left(  q,\omega\right)   &  =2\frac{\partial_{x}q\omega}{\left(  1+q\right)
}-AV\left(  x\right)  \left(  1+q\right)  .
\end{align*}
We assume that $q$ and $\omega$ belong to $B_{n,\varepsilon}^{2}$ and
$B_{0,\varepsilon}^{1}$ respectively such that%
\begin{align*}
B_{n,\varepsilon}^{2}  &  =\left\{  f\in H_{n}^{2}\left(  \Omega\right)
:f\left(  -x\right)  =f\left(  x\right)  ,\text{ }\left\Vert f\right\Vert
_{H^{2}}^{2}\leq\varepsilon^{2}\right\}  ,\\
B_{0,\varepsilon}^{1}  &  =\left\{  f\in H_{0}^{1}\left(  \Omega\right)
:f\left(  -x\right)  =f\left(  x\right)  ,\text{ }\left\Vert f\right\Vert
_{H^{1}}^{2}\leq\varepsilon^{2}\right\}  ,
\end{align*}
for\ sufficiently small $\varepsilon<<1$. Then by the Sobolev embedding
theorem (see Chapter 5 in \cite{pde}), we can easily show that%
\begin{align}
\left\Vert q\right\Vert _{\infty}^{2} \leq C\left\Vert q\right\Vert _{H^{2}%
}^{2}\leq C\varepsilon^{2}, \left\Vert \nabla q\right\Vert _{\infty}^{2} \leq
C\left\Vert q\right\Vert _{H^{2}}^{2}\leq C\varepsilon^{2}, \label{s1}%
\end{align}
for some constant $C>0.$ Therefore, for sufficiently small $\varepsilon<<1,$
$f\left(  q,\omega\right)  $ and $g\left(  q,\omega\right)  $ are well-defined
in $L^{2}\left(  \Omega\right)  .$ Moreover, we can also show that $f\left(
q,\omega\right)  \left(  -x\right)  =f\left(  q,\omega\right)  \left(
x\right)  $ and $g\left(  q,\omega\right)  \left(  -x\right)  =-g\left(
q,\omega\right)  \left(  x\right)  $ if $q$ and $\omega$ are even so that
$\int_{-1}^{1}g\left(  q,\omega\right)  =0.$

Accordingly, we can define the following operator $S\left(  q,\omega\right)
=\left(  r,w\right)  :B_{n,\varepsilon}^{2}\times B_{0,\varepsilon}%
^{1}\rightarrow H_{n}^{2}\left(  \Omega\right)  \times H_{n}^{1}\left(
\Omega\right)  ,$ where $r\ $and $w$ are solutions of (\ref{n1})-(\ref{n2})
such that they are both even. Besides, using (\ref{s1}), we can deduce that
there is a $C>0$ such that
\begin{align}
\left\Vert r\right\Vert _{H^{2}}^{2}+\left\Vert w\right\Vert _{H^{1}}^{2}  &
\leq C\left(  \left\Vert q\right\Vert _{H^{2}}^{4}+\left\Vert q\right\Vert
_{H^{2}}^{6}+\left\Vert \omega\right\Vert _{H^{1}}^{4}\left(  1+\left\Vert
q\right\Vert _{H^{2}}^{2}\right)  \right) \nonumber \\
&  +C\frac{\left\Vert q\right\Vert _{H^{2}}^{2}\left\Vert \omega\right\Vert
_{H^{1}}^{2}}{1-C\left\Vert q\right\Vert _{H^{2}}^{2}}+CA^{2}\left(
1+\left\Vert q\right\Vert _{H^{2}}^{2}\right)  ,\nonumber\\
&  \leq\varepsilon^{2},\nonumber
\end{align}
with sufficiently small $A<<1\ $so that $S:B_{n,\varepsilon}^{2}\times
B_{0,\varepsilon}^{1}\rightarrow B_{n,\varepsilon}^{2}\times B_{0,\varepsilon
}^{1}$ for sufficiently small $\varepsilon$ and $A<<1.$ Using this fact and
(\ref{s1}) again, we can estimate the differences $\widetilde{r}=r_{1}-r_{2},$
$\widetilde{w}=w_{1}-w_{2}$ for $\left(  r_{1},w_{1}\right)  =S\left(
q_{1},\omega_{1}\right)  ,$ $\left(  r_{2},w_{2}\right)  =S\left(
q_{2},\omega_{2}\right)  \in B_{n,\varepsilon}^{2} $ as
\begin{align*}
\left\Vert \widetilde{r}\right\Vert _{H^{2}}^{2}+\left\Vert \widetilde
{w}\right\Vert _{H^{1}}^{2}  &  \leq C\left(  \varepsilon\left\Vert
\widetilde{q}\right\Vert _{H^{2}}^{2}+\varepsilon\left\Vert \widetilde{\omega
}\right\Vert _{H^{2}}^{2}+A^{2}\left\Vert \widetilde{q}\right\Vert \right) \\
&  \leq\frac{1}{2}\left(  \left\Vert \widetilde{q}\right\Vert _{H^{2}}%
^{2}+\left\Vert \widetilde{\omega}\right\Vert _{H^{2}}^{2}\right)  ,
\end{align*}
for some $C>0$ and sufficiently small $\epsilon<<1.$ Then\ we can use the
Banach fixed point theorem (see Chapter 9 in \cite{pde}) so that we can show
the existence and uniqueness of solution $r$ and $w$ of (\ref{m3-1}%
)-(\ref{m3-2}).

Finally, the linearized equation of (\ref{m3-1}) on $r$ is%
\begin{equation}
r_{t}-\Delta r+r\left\vert \nabla\theta\right\vert ^{2}+2r=0. \label{rt}%
\end{equation}
Then multiplying $r$ to (\ref{rt}) and integrating in $x$ yield
\begin{equation}
\frac{d}{dt}\left\Vert r\right\Vert _{2}^{2}\leq2\left\Vert r\right\Vert
_{2}^{2}. \label{lis}%
\end{equation}
Therefore, we can conclude that $r$ is linearly stable due to (\ref{lis}). It
is known that any nonconstant steady-state of (\ref{main1}) for $A=0$ is
linearly unstable in a convex bounded domain \cite{jimbo1994stability}$.$
Moreover, the zero solution is unstable for $A=0.$ Let $\left\{
P_{0,l},\lambda_{0,l}\right\}  _{l\in S}$ be a set of linearly unstable
steady-states $\left\{  P_{0,l}\right\}  _{l\in S}$ and the first eigenvalues
$\left\{  \lambda_{0,l}\right\}  _{l\in S}$ such that $\left\{  \lambda
_{0,l}\right\}  _{l\in S}\rightarrow0$ of (\ref{main1}) for $A=0$ with an
index set $S.$ Because of the uniform boundedness of $\left\{  P_{0,l}%
\right\}  _{l\in S}$ in $H^{1}\left(  \Omega\right)  ,$ we can choose a
subsequence $\left\{  P_{0,m\left(  \sigma\right)  },\lambda_{0,m\left(
\sigma\right)  }\right\}  _{\sigma=1}^{\infty}$ of $\left\{  P_{0,l}%
,\lambda_{0,l}\right\}  _{l\in S}$ such that $V\in H_{n}^{2}\left(
\Omega\right)  $ exists such that
\begin{align*}
P_{0,m\left(  \sigma\right)  }  &  \rightarrow V\text{ weakly in }H_{n}%
^{1}\left(  \Omega\right)  ,\\
P_{0,m\left(  \sigma\right)  }  &  \rightarrow V\text{ strongly in }%
L^{2}\left(  \Omega\right)  ,
\end{align*}
as $\sigma\rightarrow0$ (See Appendix D in \cite{pde}). Then $V$ is a
steady-state, and it is not linearly unstable $\left(  \lim_{\sigma
\rightarrow\infty}\lambda_{0,m\left(  \sigma\right)  }=0\right)  $, so
$\left\vert V\right\vert \equiv1,$ but it is a contradiction because of the
definition of the stability. To conclude, there is some constant $C^{\ast}>0$
such that
\begin{equation}
0<C^{\ast}<\lambda_{0,l}, \nonumber%
\end{equation}
uniformly in $l.$

Therefore, we can complete the proof of Theorem \ref{t1} $\left(  a\right)  .$

\subsection*{\label{ic}Proof of Theorem \ref{t1} $(b)$}

We consider (\ref{m2-1})-(\ref{m2-2}) without $t.$ Moreover, we use $^{\prime
}=\frac{d}{dx}.$ Multiplying $R\left(  x\right)  \theta\left(  x\right)  $ to
(\ref{m2-2}) and integrating in $x$ yield%
\begin{align}
\int_{-1}^{1}\left(  R^{2}\theta\left(  \theta^{\prime\prime}\right)
+2R\left(  R^{\prime}\right)  \theta\left(  \theta^{\prime}\right)  \right)
dx  &  =-\int_{-1}^{1}R^{2}\left\vert \theta^{\prime}\right\vert ^{2}dx\text{
by integration by parts}\label{bbd}\\
&  =\int_{-1}^{1}AVR^{2}\theta dx\leq0,\nonumber
\end{align}
due to the Neumann boundary condition. Besides, multiplying $R$ to
(\ref{m2-2}), integrating in $x,$ and applying (\ref{bbd}), we can obtain
\begin{equation}
\left\Vert R^{\prime}\right\Vert _{2}^{2}+\int_{-1}^{1}R^{4}dx-\int_{-1}%
^{1}R^{2}\left(  1+A\theta V\right)  dx=0. \label{sec:ma-8}%
\end{equation}

Now we prove that if $V$ satisfies the conditions in Theorem \ref{t1}, then
there is no $R\left(  x\right)  $ such that $R^{\prime}\left(  x\right)
\leq0$ in $\left[  -1,0\right]  $ and $R\left(  x\right)  \geq0$ except
$R\left(  x\right)  \equiv0$ for sufficiently large $A.$ By (\ref{sec:ma-8}),
it suffices to show
\begin{equation}
-\int_{-1}^{1}R^{2}\left(  1+A\theta V\right)  dx\geq0, \label{rr}%
\end{equation}
for sufficiently large $A>0.$

We can generally assume that $R\left(  x\right)  >0$ for $-1\leq x\leq1$ if
$R\left(  x\right)  $ is nonconstant. Indeed, recall that we consider
$0<R\left(  x\right)  $ such that $R^{\prime}\left(  x\right)  \leq0$ in
$-1\leq x<0$ and $R^{\prime}\left(  x\right)  \geq0$ in $-1\leq x<0.$ If
$R\left(  -1\right)  =R\left(  1\right)  >0$ initially, then $\theta^{\prime
}\left(  0\right)  <0\ $by (\ref{m3-3}). If $R\left(  0\right)  =0$
($\min_{-1\leq x\leq1}R\left(  x\right)  =R\left(  0\right)  $), then
(\ref{m2-2}) leads to $R^{\prime}\left(  0\right)  \theta^{\prime}\left(
0\right)  =0.$ Due to $\theta^{\prime}\left(  0\right)  <0,$ $R^{\prime
}\left(  0\right)  =0.$ With the help of ODE theory, however, then we can
conclude that $R\left(  x\right)  \equiv0$ for $-1\leq x\leq1$ if $R\left(
0\right)  =R^{\prime}\left(  0\right)  =0.$

By (\ref{m3-3}) and the conditions for $V\left(  x\right)  ,$ $\theta^{\prime}
$ is nonpositive. Besides, since $R\left(  x\right)  $ attains a minimum at
$x=0$, so it leads to $\frac{R^{2}\left(  x\right)  }{R^{2}\left(  0\right)
}\geq1$ for any $-1\leq x\leq1$ and
\[
\left\vert \theta^{\prime}\left(  0\right)  \right\vert \geq-A\int_{-1}%
^{0}V\left(  s\right)  ds,
\]
by (\ref{m3-3}). Since $\theta^{\prime}\left(  x\right)  $ is continuous in
$-1<x<0$ by (\ref{m3-3}), there is a region $\widetilde{x}<x<0$ such that
\[
\left\vert \theta^{\prime}\left(  x\right)  \right\vert \geq-\frac{A}{2}%
\int_{-1}^{0}V\left(  s\right)  ds.
\]
Notice that $\widetilde{x}$ is independent of $A.$ Therefore, we obtain
\begin{align}
\theta\left(  x\right)   &  =-\int_{x}^{\widetilde{x}}\theta^{\prime}\left(
y\right)  dy-\int_{\widetilde{x}}^{0}\theta^{\prime}\left(  y\right)  dy
\geq-\int_{\widetilde{x}}^{0}\theta^{\prime}\left(  y\right)  dy\geq
\frac{A\widetilde{x}}{2}\int_{-1}^{0}V\left(  s\right)  ds,\nonumber
\end{align}
\label{theta} for $-1\leq x\leq\widetilde{x}<0.$

Using (\ref{theta}), we can compute (\ref{rr}) as%
\begin{align}
-\int_{-1}^{1}R^{2}\left(  1+A\theta V\right)  dx  &  \geq-2\int
_{-1}^{\widetilde{x}}R^{2}A\theta Vdx-2\int_{-1}^{0}R^{2}dx\label{rrr}\\
&  \geq-A^{2}\widetilde{x}\left(  \int_{-1}^{0}V\left(  s\right)  ds\right)
\int_{-1}^{\widetilde{x}}R^{2}\left(  x\right)  V\left(  x\right)
dx\nonumber\\
&  -2\int_{-1}^{\widetilde{x}}R^{2}dx-2\int_{\widetilde{x}}^{0}R^{2}%
dx.\nonumber
\end{align}
Because $0<\left\vert V\left(  x\right)  \right\vert <C$ in $-1\leq
x\leq\widetilde{x}<0$ for a constant $C>0,$ then it is easy to show
\begin{equation}
-\frac{A^{2}\widetilde{x}}{2}\left(  \int_{-1}^{0}V\left(  s\right)
ds\right)  \int_{-1}^{\widetilde{x}}R^{2}\left(  x\right)  V\left(  x\right)
dx>2\int_{-1}^{\widetilde{x}}R^{2}\left(  x\right)  dx, \label{c1}%
\end{equation}
if we choose large $A>0$. Besides, since%
\begin{align*}
-\frac{A^{2}\widetilde{x}}{2}\left(  \int_{-1}^{0}V\left(  s\right)
ds\right)  \int_{-1}^{\widetilde{x}}R^{2}\left(  x\right)  V\left(  x\right)
dx  &  \geq-\frac{A^{2}\widetilde{x}}{2}\left(  \int_{-1}^{\widetilde{x}%
}V\left(  s\right)  ds\right)  ^{2}R^{2}\left(  \widetilde{x}\right)  ,\\
-2\widetilde{x}R^{2}\left(  \widetilde{x}\right)   &  \geq2\int_{\widetilde
{x}}^{0}R^{2}dx,
\end{align*}
if we choose $A>0$ such that%
\[
-\frac{A^{2}\widetilde{x}}{2}\left(  \int_{-1}^{\widetilde{x}}V\left(
s\right)  ds\right)  ^{2}R^{2}\left(  \widetilde{x}\right)  >-2\widetilde
{x}R^{2}\left(  \widetilde{x}\right)  ,
\]
then it is also obtainable that%
\begin{equation}
-\frac{A^{2}\widetilde{x}}{2}\left(  \int_{-1}^{0}V\left(  s\right)
ds\right)  \int_{-1}^{\widetilde{x}}R^{2}\left(  x\right)  V\left(  x\right)
dx>2\int_{\widetilde{x}}^{0}R^{2}dx, \label{c2}%
\end{equation}
for large $A>0.$ Consequently, substituting (\ref{c1}) and (\ref{c2}) into
(\ref{rrr}), we can show (\ref{rr}).

Hence, Theorem \ref{t1} $(b)$ follows immediately.


\begin{thebibliography}{99}                                                                                               %
\providecommand{\url}[1]{#1}

\providecommand{\newblock}{\relax} \providecommand{\bibinfo}[2]{#2}
\providecommand\BIBentrySTDinterwordspacing{\spaceskip=0pt\relax}
\providecommand\BIBentryALTinterwordstretchfactor{4}
\providecommand\BIBentryALTinterwordspacing{\spaceskip=\fontdimen2\font plus
\BIBentryALTinterwordstretchfactor\fontdimen3\font minus
\fontdimen4\font\relax}
\providecommand\BIBforeignlanguage[2]{{\expandafter\ifx\csname l@#1\endcsname\relax
\typeout{** WARNING: IEEEtran.bst: No hyphenation pattern has been}\typeout{** loaded for the language `#1'. Using the pattern for}\typeout{** the default language instead.}\else
\language=\csname l@#1\endcsname
\fi
#2}}

\bibitem {rp1}L.~Mestel, \emph{Stellar magnetism}.\hskip 1em plus 0.5em minus
0.4em\relax OUP  Oxford, 2012, vol. 154.

\bibitem {rp2}J.~D. Murray, \emph{Mathematical biology [electronic resource].:
An introduction}.\hskip 1em plus 0.5em minus 0.4em\relax Springer, 2002.

\bibitem {eb3}K.~Burrell, ``Effects of e $\times$ b velocity shear and
magnetic shear on  turbulence and transport in magnetic confinement devices,''
\emph{Physics of Plasmas}, vol.~4, no.~5, pp. 1499--1518, 1997.

\bibitem {eb5}J.~Cornelis, R.~Sporken, G.~Van~Oost, and R.~Weynants,
``Predicting the radial  electric field imposed by externally driven radial
currents in tokamaks,''  \emph{Nuclear fusion}, vol.~34, no.~2, p. 171, 1994.

\bibitem {eb2}R.~Groebner, ``An emerging understanding of h-mode discharges in
tokamaks,''  \emph{Physics of Fluids B: Plasma Physics}, vol.~5, no.~7, pp.
2343--2354,  1993.

\bibitem {eb4}R.~Taylor, M.~Brown, B.~Fried, H.~Grote, J.~Liberati,
G.~Morales, P.~Pribyl,  D.~Darrow, and M.~Ono, ``H-mode behavior induced by
cross-field currents in a  tokamak,'' \emph{Physical review letters}, vol.~63,
no.~21, p. 2365, 1989.

\bibitem {eb1}F.~Wagner, G.~Becker, K.~Behringer, D.~Campbell, A.~Eberhagen,
W.~Engelhardt,  G.~Fussmann, O.~Gehre, J.~Gernhardt, G.~v. Gierke,
\emph{et~al.}, ``Regime of  improved confinement and high beta in
neutral-beam-heated divertor discharges  of the asdex tokamak,''
\emph{Physical Review Letters}, vol.~49, no.~19, p.  1408, 1982.

\bibitem {eb6}R.~R. Weynants, G.~Van~Oost, G.~Bertschinger, J.~Boedo, P.~Brys,
T.~Delvigne,  K.~Dippel, F.~Durodie, H.~Euringer, K.~Finken, \emph{et~al.},
``Confinement  and profile changes induced by the presence of positive or
negative radial  electric fields in the edge of the textor tokamak,''
\emph{Nuclear Fusion},  vol.~32, no.~5, p. 837, 1992.

\bibitem {krebs2013nonlinear}I.~Krebs, M.~Hoelzl, K.~Lackner, and
S.~G{\"u}nter, ``Nonlinear excitation of  low-n harmonics in reduced
magnetohydrodynamic simulations of edge-localized modes,'' \emph{Physics of
Plasmas}, vol.~20, no.~8, p. 082506, 2013.

\bibitem {lee2017solitary}J.~Lee, G.~Yun, W.~Lee, M.~Kim, M.~Choi, J.~Lee,
M.~Kim, H.~Park, J.~Bak,  W.~Ko, \emph{et~al.}, ``Solitary perturbations in
the steep boundary of  magnetized toroidal plasma,'' \emph{Scientific
Reports}, vol.~7, p. 45075,  2017.

\bibitem {wenninger2012solitary}R.~Wenninger, H.~Zohm, J.~Boom, A.~Burckhart,
M.~Dunne, R.~Dux, T.~Eich,  R.~Fischer, C.~Fuchs, M.~Garcia-Munoz,
\emph{et~al.}, ``Solitary magnetic  perturbations at the elm onset,''
\emph{Nuclear Fusion}, vol.~52, no.~11, p.  114025, 2012.

\bibitem {beyer2005nonlinear}P.~Beyer, S.~Benkadda, G.~Fuhr-Chaudier,
X.~Garbet, P.~Ghendrih, and  Y.~Sarazin, ``Nonlinear dynamics of transport
barrier relaxations in tokamak  edge plasmas,'' \emph{Physical review
letters}, vol.~94, no.~10, p. 105001,  2005.

\bibitem {m1}S.-I. Itoh, K.~Itoh, A.~Fukuyama, and Y.~Miura, ``Edge localized
mode activity  as a limit cycle in tokamak plasmas,'' \emph{Physical review
letters},  vol.~67, no.~18, p. 2485, 1991.

\bibitem {leconte2016ginzburg}M.~Leconte, Y.~Jeon, and G.~Yun,
``Ginzburg-landau model in a finite  shear-layer and onset of transport
barrier nonlinear oscillations: A paradigm  for typeiii elms,''
\emph{Contributions to Plasma Physics}, vol.~56, no. 6-8,  pp. 736--741, 2016.

\bibitem {m2}J.~L{\"o}nnroth, V.~Parail, G.~Corrigan, D.~Heading, G.~Huysmans,
A.~Loarte,  S.~Saarelma, G.~Saibene, S.~Sharapov, J.~Spence, \emph{et~al.},
``Integrated  predictive modelling of the effect of neutral gas puffing in
elmy h-mode  plasmas,'' \emph{Plasma physics and controlled fusion}, vol.~45,
no.~9, p.  1689, 2003.

\bibitem {ls}J.~Connor, R.~Hastie, H.~Wilson, and R.~Miller,
``Magnetohydrodynamic stability  of tokamak edge plasmas,'' \emph{Physics of
Plasmas}, vol.~5, no.~7, pp.  2687--2700, 1998.

\bibitem {r1}I.~S. Aranson and L.~Kramer, ``The world of the complex
ginzburg-landau  equation,'' \emph{Reviews of Modern Physics}, vol.~74, no.~1,
p.~99, 2002.

\bibitem {jimbo1994stability}S.~Jimbo and Y.~Morita, ``Stability of
nonconstant steady-state solutions to a  ginzburg--landau equation in higher
space dimensions,'' \emph{Nonlinear Analysis: Theory, Methods \&
Applications}, vol.~22, no.~6, pp. 753--770,  1994.

\bibitem {li2014global}F.~Li and B.~You, ``Global attractors for the complex
ginzburg--landau  equation,'' \emph{Journal of Mathematical Analysis and
Applications}, vol.  415, no.~1, pp. 14--24, 2014.

\bibitem {w5}H.~Lu, P.~W. Bates, S.~L{\"u}, and M.~Zhang, ``Dynamics of the
3-d fractional  complex ginzburg--landau equation,'' \emph{Journal of
Differential Equations}, vol. 259, no.~10, pp. 5276--5301, 2015.

\bibitem {yun2011two}G.~Yun, W.~Lee, M.~Choi, J.~Lee, H.~Park, B.~Tobias,
C.~Domier, N.~Luhmann~Jr,  A.~Donn{\'e}, J.~Lee, \emph{et~al.},
``Two-dimensional visualization of  growth and burst of the edge-localized
filaments in kstar h-mode plasmas,''  \emph{Physical review letters}, vol.
107, no.~4, p. 045004, 2011.

\bibitem {mp3}J.~Lee, G.~S. Yun, M.~J. Choi, J.-M. Kwon, Y.-M. Jeon, W.~Lee,
N.~C.  Luhmann~Jr, and H.~K. Park, ``Nonlinear interaction of edge-localized
modes  and turbulent eddies in toroidal plasma under n= 1 magnetic
perturbation,''  \emph{Physical Review Letters}, vol. 117, no.~7, p. 075001, 2016.

\bibitem {pde}L.~Evans, \emph{Partial Differential Equations}, ser. Graduate
Studies in  Mathematics.\hskip 1em plus 0.5em minus 0.4em\relax American
Mathematical  Society, Providence, 2010.
\end{thebibliography}
\end{document}